\documentclass[conference]{IEEEtran}
\IEEEoverridecommandlockouts

\usepackage{cite}
\usepackage{amsmath,amssymb,amsfonts}
\usepackage{amsmath,amssymb}
\usepackage{centernot}

\usepackage{graphicx}
\usepackage{textcomp}
\usepackage{xcolor}
\usepackage{booktabs} 
\usepackage{graphicx}
\graphicspath{ {figures/} }
\graphicspath{ {figures2/} }
\usepackage{algpseudocode}
\usepackage{algorithm}
\usepackage[T1]{fontenc}

\usepackage{caption}

\usepackage{url}

\usepackage{booktabs}
\usepackage{balance}

\def\BibTeX{{\rm B\kern-.05em{\sc i\kern-.025em b}\kern-.08em
    T\kern-.1667em\lower.7ex\hbox{E}\kern-.125emX}}
\begin{document}

\title{COVID-bit: Keep a Distance of (at least) 2m From My Air-Gap Computer!}

\author{\IEEEauthorblockN{Mordechai Guri}
	\IEEEauthorblockA{Ben-Gurion University of the Negev, Israel\\
		Department of Software and Information Systems Engineering\\ Cyber-Security Research Center \\
		Email: gurim@post.bgu.ac.il \\ Demo video: http://www.convertchannels.com}}

\maketitle

\begin{abstract}
Air-gapped systems are isolated from the Internet due to the sensitive information they handle. 

This paper presents COVID-bit, a new \textbf{COV}ert channel attack that leaks sensitive information over the air from highly isolated systems. The information emanates from the air-gapped computer over the air to a distance of 2m and more and can be picked up by a nearby insider or spy with a mobile phone or laptop. Malware on an air-gapped computer can generate radio waves by executing crafted code on the target system. The malicious code exploits the dynamic power consumption of modern computers and manipulates the momentary loads on CPU cores. This technique allows the malware to control the computer's internal utilization and generate low-frequency electromagnetic radiation in the 0 - 60 kHz band. Sensitive information (e.g., files, encryption keys, biometric data, and keylogging) can be modulated over the emanated signals and received by a nearby mobile phone at a max speed of 1000 bits/sec. We show that a smartphone or laptop with a small \$1 antenna carried by a malicious insider or visitor can be used as a covert receiver. Notably, the attack is highly evasive since it executes from an ordinary user-level process, does not require root privileges, and is effective even within a Virtual Machine (VM). We discuss the attack model and provide technical details. We implement air-gap transmission of texts and files, and present signal generation and data modulation. We test the covert channel and show evaluation results. Finally, we present a set of countermeasures to this air-gap attack. 
\end{abstract}

\begin{IEEEkeywords}
	air-gap, network, exfiltration, electromagnetic, leakage, covert channel 
\end{IEEEkeywords}

\section{Introduction}
An `air-gap' is a security measure that describes the physical isolation of a network or device from external networks such as the Internet. In certain cases, the term is extended to describe isolation from \textit{any} less-secure networks, public or non-public. This measure is taken especially when sensitive or highly confidential data is involved, to minimize data leakage risks. 

Various defensive policies enforce air-gap isolation. For example, air-gapped computers usually have their wireless interface controllers (E.g., Wi-Fi and Bluetooth) disabled or removed. In some cases, physical access to the air-gapped environment is regulated via strict access control, biometric authentication, and personal identity systems. 

Military organizations commonly manage air-gapped networks to maintain classified information. For example, the United States Department of Defense uses the Secret Internet Protocol Router Network (SIPRNet) to transmit classified information to the level of Secret \cite{dinsmore2016niprnet}. The National Security Agency Network (NSANET) is the official NSA intranet that handles information up to the level of Top-Secret (TS/SCI) \cite{Electros46:online}. Other types of facilities may also be maintained within air-gapped environments, including banking and stock exchange networks, critical infrastructure, life-critical systems, and medical equipment. 

\subsection{Air-Gap Penetration}
But despite the high level of isolation, air-gapped networks are not invulnerable to cyber-attacks. Over the years, it has been shown that motivated adversaries can hack even highly-secured air-gapped networks. To achieve this goal, the attackers may use various strategies such as interfering with the supply chains, employing social engineering techniques, and using malicious insiders. One of the famous breaches of an air-gapped network was revealed in November 2008 and was described as ``the most serious breach of the U.S. military's classified computer systems.'' \cite{AgentBTZ65:online}. In this case, sophisticated malware is known as Agent.Btz is delivered to the strategic network via a removable USB drive and spreads in the internal networks in a computer worm. The Stuxnet worm, discovered in 2011, is 500-kilobyte malware that attacked industrial networks, including a uranium enrichment plant in Iran. Stuxnet could spread stealthily between computers, even those not connected to the Internet \cite{kushner2013real}. Over the past decade, several publicized attack incidents on air-gapped networks have been reported. In 2020 ESET researchers discovered Ramsay, a cyber-espionage framework capable of operating within air-gapped networks and tailored for collecting and exfiltration of sensitive documents \cite{ESETRese49:online}. In 2021, ESET researchers published technical information on various APTs used to attack air-gapped networks in the last 15 years \cite{dorais2021jumping}.

\subsection{Air-Gap Exfiltration}
After breaching the air gap, spreading over the internal network, and gathering sensitive information, the adversary has to leak the collected data outward. While this exfiltration is a trivial task in the case of Internet-connected networks, it is a much more challenging task in the case of isolated, air-gapped networks. Due to the \textit{lack of connectivity} to the Internet, the attacker must resort to special communication techniques, also known as `air-gap covert channels.' In these techniques, the attacker exploits non-standard, physical mediums to leak information outside the isolated environment. Over the years, various types of air-gap covert channels have been explored by researchers, mainly with electromagnetic, acoustic, and optical mediums. With the electromagnetic methods, attackers use electromagnetic waves emanating from the computer parts to encode and leak sensitive information \cite{Kuhn2002}. An attacker can also use ultrasound and lights to exfiltrate data out of the air-gapped systems \cite{wong2018crossing}. These innovative techniques enable adversaries to deliver sensitive data from isolated, air-gapped systems.

\subsection{Contribution}
This paper presents a new air-gap covert channel that enables attackers to leak data from air-gapped systems. The contributions of our work are outlined below.

\begin{enumerate}
	
	\item \textbf{Attack model.} We present an adversarial attack model for isolated, air-gapped networks. We also discuss the bifurcated attack, in which the transmitter and receiver maintain covert communication.

	\item \textbf{Technique.} Our technique is based on the switching mode power supply and works on a wide range of computers and devices.

	\item {\textbf{Stealth.}} The technique is evasive and stealthy since it runs from standard user-mode processes and does not require high privileges.

	\item {\textbf{Implementation.}} We present the implementation of a covert receiver that can be used with any ordinary smartphone.

	\item {\textbf{Exfiltration.}} We implement various types of data exfiltration, including air-gap exfiltration of text and files.

	\item {\textbf{Defense.}} We discuss a set of countermeasures to cope with this threat.
	
\end{enumerate}

This paper is organized as follows. The adversarial attack model is explained in Section \ref{sec:attack}. We review related work in Section \ref{sec:related}. Technical background is provided in Section \ref{sec:tech}. The design and implementation of the transmitter and receiver are described in Sections \ref{sec:trans}, \ref{sec:rec}, respectively. Section \ref{sec:eval} describes the analysis and evaluation results. Countermeasures are discussed in Section \ref{sec:counter}. We conclude in Section \ref{sec:conclusion}.

\begin{figure}[]
	\centering
	\captionsetup{labelformat=empty}
	\includegraphics[width=\linewidth]{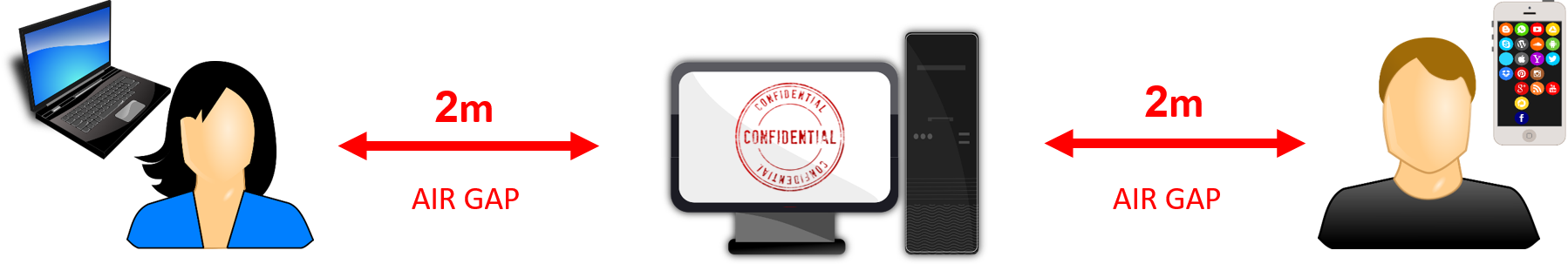}
\end{figure}

\section{Adversarial Attack Model}
\label{sec:attack}
The adversarial attack model is a bifurcated compound of two components:
\begin{enumerate}
	\item \textbf{Transmitter.} This transmitting computer(s) belongs to the air-gapped network. 
	
	\item \textbf{Receiver.} This component (e.g., smartphone) receives the exfiltrated data, decodes it, and delivers it to the attacker. 
\end{enumerate}

The chain of attack consists of three main stages.

\subsection{Air-gap Breaching}
Lockheed Martin adopted the APT kill chain framework from a military attack model on physical targets. The initial phases in the kill chain of a typical APT include reconnaissance and intrusion. In the case of a standard Internet-connected network, attackers may use attack vectors such as browser vulnerabilities and malicious email attachments to penetrate the organizational network. However, other attack methods must be adopted to attack air-gapped networks that are separated from the Internet, such as supply chain attacks, malicious insiders, and deceived employees. These attack vectors were demonstrated in several cyber incidents in the past. For example, Agent.btz was a computer worm that compromised U.S. military networks in 2008 via an infected USB drive used by a deceived employee. The Stuxnet worm attacked supervisory control and data acquisition (SCADA) systems in 2010 \cite{kushner2013real}. During 2018 and 2019, the internal networks of critical infrastructure facilities in the U.S. and India were compromised \cite{Nobigdea65:online}\cite{AnIndian12:online}. In June 2020, researchers in the Kaspersky security firm reported USBCulprit, an Advanced Persistent Threat (APT) which seems to be designed to reach air-gapped networks \cite{guri2021usbculprit}. Other malware that targets air-gap environments, such as Ramsay \cite{ESETRese49:online} and Tick \cite{TickGrou19:online}, were reported in recent years. Table \ref{tab:attacks} lists a few APT attacks on air-gapped facilities that have been reported in the past.

Note that although the term advanced persistent threat (APT) was used originally to describe nation-state cyberattacks, over the years, the term has broadened to describe a wide variety of cyber-crime attacks targeted at businesses, finance sectors, and civilian industries as well.

\begin{table}[]
	\caption{{Cyberattacks on air-gapped networks}}
	\label{tab:attacks}
	\resizebox{\columnwidth}{!}{
		\begin{tabular}{@{}ll@{}}
			\toprule
			Attack             & Target facilities                                                                                    \\ \midrule
			Agent.BTZ \cite{AgentBTZ65:online}           & U.S military classified networks                                                                     \\
			Stuxnet worm \cite{kushner2013real}        & Uranium enrichment facility                                                       \\
			Fanny worm \cite{AFannyEq68:online}         & Industrial IT networks                                                                    \\
			SymonLoader  \cite{TickGrou19:online}           & Critical infrastructure and manufacturing                                                 \\
			Copperfield \cite{AnotherC19:online}          & Critical infrastructure networks                                                          \\
			ProjectSauron \cite{TheProje50:online}       & Government, military, and telecommunication                                                                                                           \\
			USBCulprit \cite{guri2021usbculprit}       & \begin{tabular}[c]{@{}l@{}}Diplomatic and government sectors \end{tabular}   \\
			Ramsay \cite{NewRamsa19:online}             & Physically isolated systems                                                               \\ \bottomrule
	\end{tabular}}
\end{table}

\subsection{Receiver Maintaining}
There are two strategies to maintain the receiver in this attack model:

\subsubsection {Malicious insiders (an insider threat)} An insider threat is a security risk that originates from people associated with the organization, such as employees, contractors, and service providers. Another type of insider is a visitor who is technically an outsider who has access to secure facilities. In the context of the attack model, a receiver mobile phone might be intentionally carried by an employee or visitor. In this case, the insider holds the mobile phone near one or more sensitive, compromised computers. Malicious insider incidents from the last decade include the famous case of Edward Snowden, the consultant who leaked highly classified information from the National Security Agency (NSA) \cite{bauman2014after}. In October 2021, the United States Department of Justice announced a leakage of classified information related to the design of the nuclear-powered Virginia-class submarine. According to the reports, the data was stolen by a worker, and was leaked on an SD card hidden in a peanut butter sandwich \cite{Submarin22:online}.

\subsubsection {Implanted receivers \& Supply chain attacks} The attacker may hide tiny receivers nearby the air-gapped computers. Such operations can be carried out by malicious insiders or via supply chains. Notably, tampering with the supply chains is shown to be feasible by capable and skilled adversaries \cite{reed2014supply}\cite{gupta2020additive}. A typical supply chain attack involves physically tampering with computing systems to deliver malware into the organization. For example, the leaked National Security Agency (NSA) catalog published by the German news magazine Der Spiegel in December 2013 includes components such as USB hardware that provides covert remote access to the target machine \cite{COTTONMO81:online}. In October 2018, Bloomberg Businessweek claimed that China used a tiny implanted hardware chip to infiltrate America's top companies. According to the report, investigators found that the supply chain attack affected almost 30 companies, including a major bank, government contractors, and the world's most valuable company, Apple Inc.

\subsection{Exfiltration}
In the exfiltration stage, the malware transmits the data using covert signals generated from the computers via the COVID-bit covert channel. The data can be documents, keystroke logging, biometric identifiers, access credentials, encryption keys, or any information valuable to the attacker. A nearby mobile phone or implanted receiver detects the transmission, decodes the data, and delivers it to the attacker via the Internet.

\begin{figure}
	\centering
	\includegraphics[width=\linewidth]{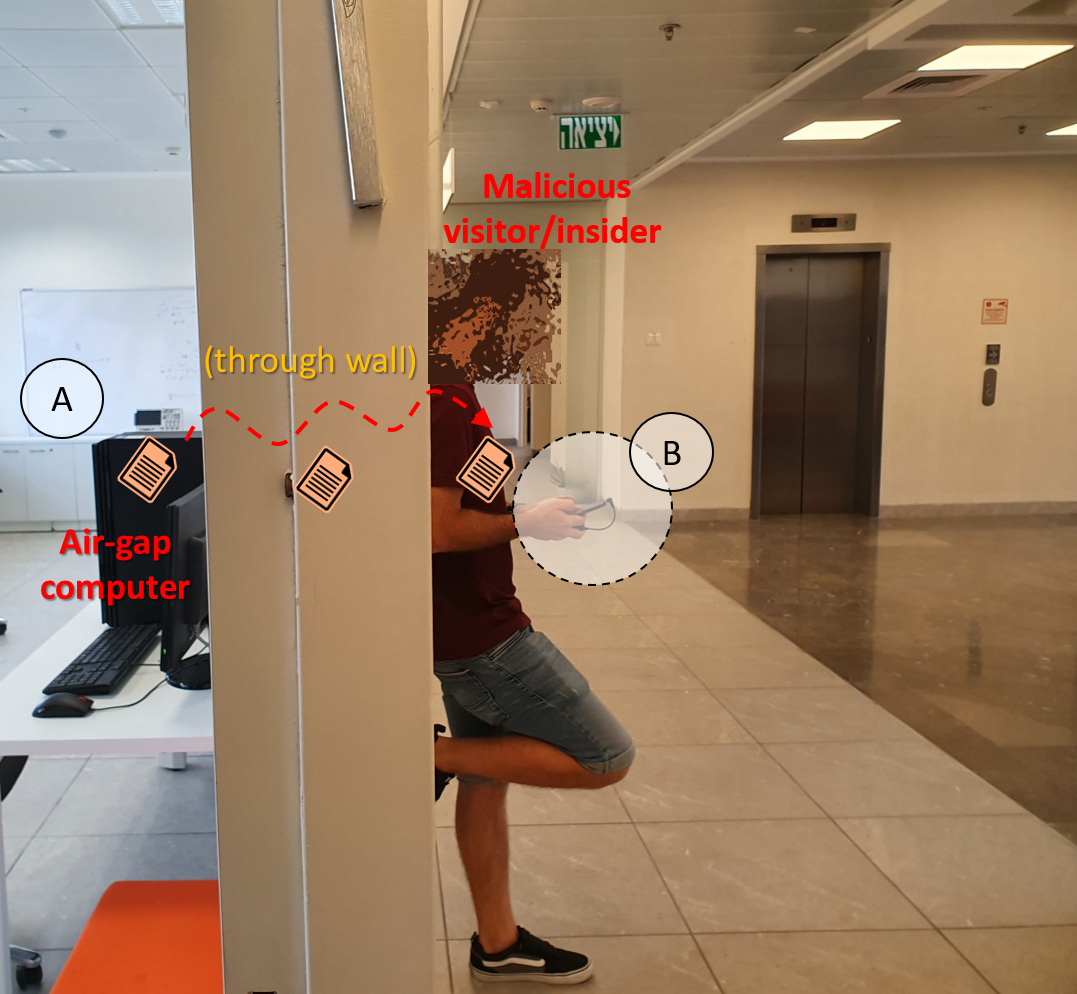}
	\caption{Attack scenario (`insider'): malware within the infected air-gapped computer (A) transmits the sensitive file via electromagnetic radiation. The file is received by a smartphone of a malicious insider/visitor standing behind the wall in a less secure area (B).}
	\label{fig:scenario}
\end{figure}

\subsection{Transmitter-Receiver Attack Model}
The attack is illustrated in Figure \ref{fig:scenario}. Malware within the infected air-gapped computer (A) transmits sensitive file. The file is received by the smartphone of a malicious insider/visitor located behind the wall in a less secure area (B). 

It is important to note that the attack model compound of transmitters and receivers has been broadly discussed in the domain of covert channels for the last twenty years. In this attack model, the attacker must compromise the targeted system and infect or maintain electronic devices such as radio receivers, microphones, and optical cameras. In reward for their effort, the attackers may collect valuable information from air-gapped networks. Note that many covert channel studies in recent years share the transmitter-receiver attack model, including AirHopper \cite{guri2014airhopper}, Jabber \cite{9300268}, LORAEMR \cite{LORAMER}, ODINI \cite{guri2019odini}, Nicscatter \cite{yang2017nicscatter}, and MagView \cite{zhang2020magview}.

\section{Related Work}
\label{sec:related}
Air-gap covert channels are communication channels that enable attackers to leak data from isolated, air-gapped systems in non-conventional ways. 
These methods can be mainly classified into electromagnetic, magnetic,  electric, optical, thermal, and acoustic.

With electromagnetic covert channels, the malicious code in the system exploits the electromagnetic radiation from components to modulate and transmit data outward. Electromagnetic-based channels have been commonly explored in this field for many years. Kuhn \cite{kuhn1998soft} introduced the use of video cards to broadcast radio signals modulated with information from workstations. In the context of air-gap attacks, AirHopper attack \cite{guri2014airhopper} used the video cards in air-gapped computers to generate FM signals modulated with the leaked information. Similarly, GSMem \cite{guri2015gsmem}, USBee \cite{guri2016usbee}, BitJabber \cite{9300268}, EMR \cite{LORAMER}, Air-Fi \cite{guri2020air}, and CloakLoRa \cite{9259364} introduced attack scenarios in which attackers use different sources of radiation from air-gapped systems for data exfiltration. Note that electromagnetic radiation can be prevented with Faraday cages. Researchers proposed using the magnetic fields for covert communication to leak data from Faraday caged air-gapped computers to nearby magnetic sensors \cite{guri2019odini, ,GURI2021115}. 
Electricity can also be used to leak data. In 2019, researchers introduced `PowerHammer' - an attack that enables data exfiltration via emission conducted to the power lines \cite{guri2019powerhammer}. In 2020, Shao et al. exploited the noise induced by the power lines for data transmission \cite{shao2020your}. The optical medium can also be used for air-gap communication. In this attack, an optical emanating source is used to modulate signals, which are received by remote cameras or optical sensors. In this context, PC keyboard LEDs were studied by Loughry \cite{loughry2002information} and more recently by Guri \cite{guri2019ctrl}. Nassi et al. established a covert channel through the air-gap using malware, remote lasers, and scanners \cite{nassi2018xerox}. In the same manner, the researcher proposed using HDD, network devices, and camera LEDs for data leakage \cite{Guri2017}. Computer screens can also be used to leak data. Researchers used the screen brightness \cite{guri2019brightness}, and hidden images \cite{guri2016optical} to exfiltrate data in stealthy ways. Thermal-based covert channels such as BitWhisper and HOTSPOT allow maintaining a special type of covert communication between two air-gapped computers via the manipulation of temperature \cite{guri2015bitwhisper}. In acoustic covert channels, data is transmitted via sound waves in various frequencies. Madhavapeddy \cite{madhavapeddy2005audio} discussed audio-based networking between computers via standard speakers and a built-in microphone. Hanspach et al. \cite{hanspach2014covert} extended this method for malicious purposes where the attacker uses inaudible frequencies to establish covert mesh networking between laptops in a room. Acoustic covert communication without microphones is also possible, as shown in MOSQUITO \cite{Guri2018Mosquito}, a covert communication channel between two air-gapped computers via ultrasonic communication between speakers. 
Researchers presented air-gap covert channels for highly secured environments where speakers are not present. They used various components to leak data acoustically, including computer fans \cite{guri2020fansmitter}, and HDD drives \cite{Guri2017f}.

\section{Technical Background}
\label{sec:tech}
Computers consume power via their power supplies, which transfer power from a DC or AC source to a DC load. In switch-mode power supplies (SMPS), voltage regulation is made by a continuous change in the ratio of on-to-off time, also known as duty cycles. The DC supply from a rectifier or battery is fed to the inverter, turning it on and off at high rates by switching MOSFET or power transistors. This switching rate is also known as the \textit{switching frequency}. Due to their high efficiency, SMPSs are used today in all types of electronic equipment, including PC workstations, laptops, the Internet of Things (IoT), vehicles, and mobile devices. An in-depth discussion on the internal design of SMPS can be found in relevant literature \cite{handbookbillings2011switchmode}. 

\subsection{Dynamic Frequency}

The two main types of electromagnetic emissions from SMPSs

are conducted and radiated. The conducted emissions are carried by wires such as power cables, signal cables, and internal buses. The radiated emissions propagate in the form of electromagnetic fields. 

Part of the radiated emission of SMPS is related to currents being switched. More specifically, the waveforms generated from this emission correlate with the system's momentary switching frequency. The switching frequency in modern systems is changed during the execution due to the dynamic voltage and frequency scaling mechanism \cite{le2010dynamic}. Dynamic voltage and frequency scaling is used in all modern architectures to allow the system to adjust the frequency and supply voltage to a specific load in the computer. This mechanism significantly improves the processor's efficiency by running at lower voltages when possible. Since modern systems are energy efficient and use dynamic scaling, the workload of the CPU directly affects the dynamic changes in its power consumption \cite{von2016variations}. This is maintained by the voltage regulator module in the system. 

Consequently, the speed at which a circuit operates is proportional to the voltage differential in that circuit and the switching frequency. By regulating the workload of the CPU, it is possible to govern its power consumption and hence control the momentary switching frequency of the SMPS. The electromagnetic radiation generated by this intentional process can be received from a distance using appropriate antennas. In the technique used in the COVID-bit attack, we control the switching frequency to generate electromagnetic radiation in the low-frequency band (0 - 60 kHz).

\section{Transmitter Implementation}
\label{sec:trans}

\begin{figure}
	\centering
	\includegraphics[width=\linewidth]{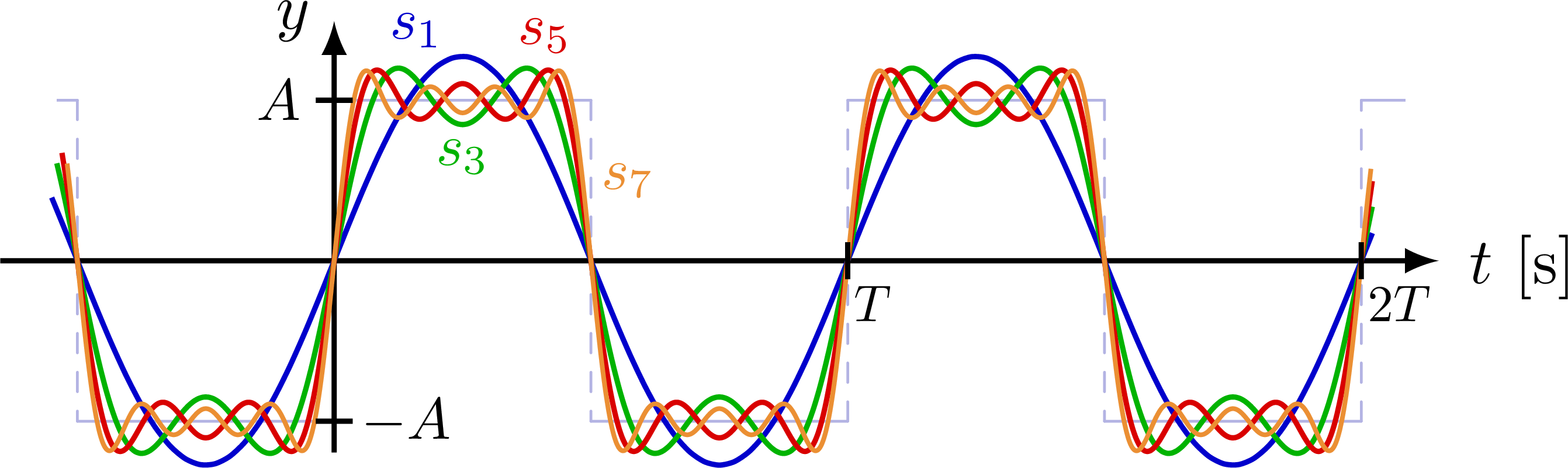}
	\caption{A square wave and corresponding Furrier series s1, s2,..., s7.}
	\label{fig:f5}
\end{figure}

As noted, the primary source of electromagnetic radiation in SMPS is because of their internal design and switching characteristics. In the conversion from AC-DC and DC-DC, the MOSFET switching components turning on or off at specific frequencies create a square wave. Fourier series can represent this square wave as the summation of sine waves with harmonically-related frequencies. Equation \ref{eqn1}) shows the Fourier series of square wave with an amplitude of 1 having a cycle frequency $f$ over time $t$. Figure \ref{fig:f5} illustrates the Furrier expansion of a square wave in the time domain.
The full Fourier spectrum of harmonics becomes electromagnetic radiation resulting from the switching action.

\begin{equation} \label{eqn1}
f(x)=\frac{4}{\pi}\sum_{n=1}^{\infty}\frac{sin(2\pi(2n-1)ft)}{2n-1}
\end{equation}

The second source of radiation is the fast current (dI/dt) and voltage (dV/dt) transitions. Due to maxwell’s equation, this alternating currents and voltages produce an alternating electromagnetic field with a magnitude that reduces with distance. Some copper components in the system act like antennas and cause radiation at correlated frequencies.

\subsection{Switching Control}
Due to the dynamic power regulation, the momentary load on a specific core is correlated with the voltage it consumes. This is achieved by a component called the voltage regulator module (VRM), also referred to as the processor power module (PPM). For example, Intel Haswell CPUs feature voltage-regulation components on the same package (or die) as the CPU. To generate emission in a frequency $f_c$, we continuously apply a workload on a specific core in a frequency $f_c$. From the operating system's perspective, signal generation is done by starting and stopping CPU-bound execution code at frequency $f_c$. Our signal generation approach used several cores concurrently to generate more substantial emissions (since more cores consume more power). A governor thread synchronizes between several worker threads in the system, whereas each thread is bound to a specific core. Algorithms 1 - 2 show the pseudo-code for the governor thread, and synchronization handler, respectively, and outline the signal generation process.

\begin{figure}[]
	\centering
	\captionsetup{labelformat=empty}
	\includegraphics[width=\linewidth]{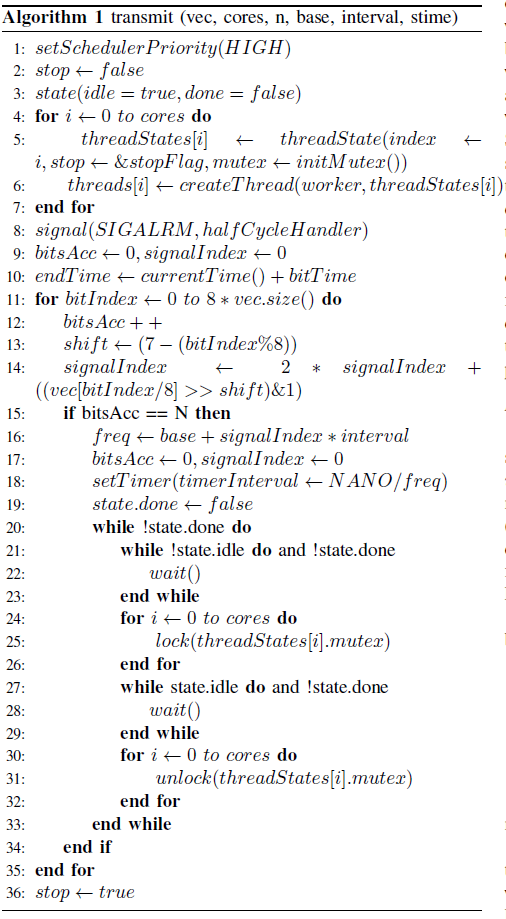}
	\includegraphics[width=\linewidth]{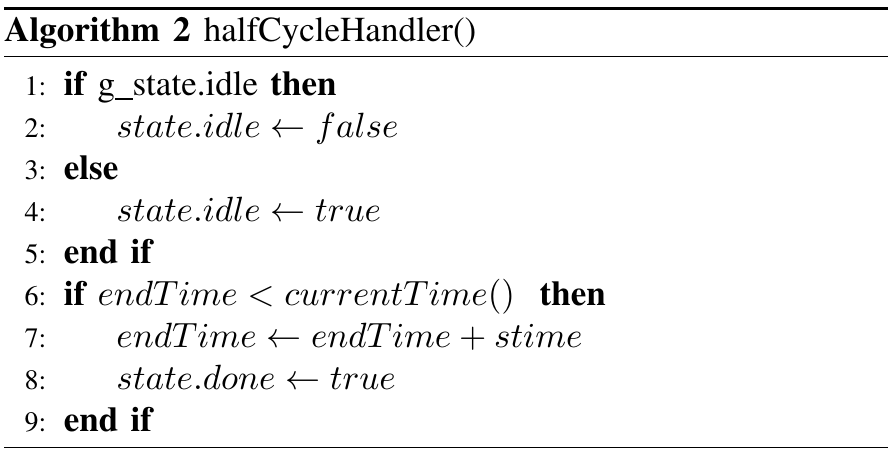}
\end{figure}

The function \textit{transmit} is called to transmit $n$ bits from the $vec$ array. The main thread is responsible for generating the emission ($base$) for a time duration of ($interval$) milliseconds, using ($cores$) cores. 
We create $cores$ worker threads with corresponding synchronization flags and Mutex. We also bind each to a specific logical core. To generate the carrier wave, each thread strains its core at a frequency $f_c$. This is accomplished by repeatedly alternating between a continuous workload and idle state for time periods of $1/2f_c$. 
We set a SIGALRM to the half-cycle handler function (line 8). This signal is raised when a time interval specified in a call to the alarm function expires. 
The timer cycle in nanoseconds is calculated from the frequency of the signal (line 18). Finally, the main thread governs the synchronization flags for the corresponding cycle times. This is done for a time period of the signal to lock and unlock the synchronization object for the worker threads. This ensures that the generation of duty cycles is precisely timed between the cores. Based on the algorithms described above, we implement a transmitter program for Linux Ubuntu OS.

\subsection{Data Transfer}
\label{datatransfer}

Using the method described above, we could generate a signal at a specific frequency $f_c$ for a time duration of $interval$. The binary data was transmitted using the fundamental frequency-based modulation. In frequency-shift keying (FSK), the data is represented by a specific frequency of a carrier wave. In the binary FSK (B-FSK), each of the two frequencies represents a different symbol (`1' or `0'). A B-FSK signal representation is shown in equation \ref{eqn2} for binary 1,' and 0', where $E_{b}$ is the transmitted signal energy, $T_{b}$ is the bit period, and $\phi_{1}$, $\phi_{2}$ are initial phases at $t$ = 0.

\begin{alignat}{1}
\label{eqn2}
s_{1}(t))=\sqrt{\frac{2E_{b}}{T_{b}}}cos(2\pi f_{1}t+\phi_{1} ), 0\leq t\leq T_{b} \\    
s_{2}(t))=\sqrt{\frac{2E_{b}}{T_{b}}}cos(2\pi f_{2}t+\phi_{2} ), 0\leq t\leq T_{b} \nonumber
\end{alignat}

In our case, two different switching frequencies are used for modulation.

Since we have multiple cores, we could use different cores to generate signals in different frequencies. The straightforward approach is to utilize M cores for M-FSK modulation. Equation \ref{eqn3} shows the M-FSK, which is the generalization of the B-FSK case, where $E_{b}$ is the transmitted signal energy, $T_{b}$ is the bit period, and $\phi_{m}$ is the initial phases at $t$ = 0.

\begin{alignat}{1}
 	\label{eqn3}
 	s_{m}(t))=\sqrt{\frac{2E_{b}}{T_{b}}}cos(2\pi f_{m}t+\phi_{m} ), 0\leq t\leq T_{b}     
 \end{alignat}

It is also possible to use OFDM (Orthogonal frequency division multiplexing). OFDM is a digital multi-carrier modulation scheme that extends single-carrier modulation. Instead of modulating binary data with a single sub-carrier, we employ several sub-carriers concurrently. With this modulation technique, bandwidth be utilized efficiently and advanced channel coding can be used.

By governing the utilization of each core separately, we could use a different sub-carrier for each transmitting core. 

\begin{figure}
	\centering
	\includegraphics[width=1\linewidth]{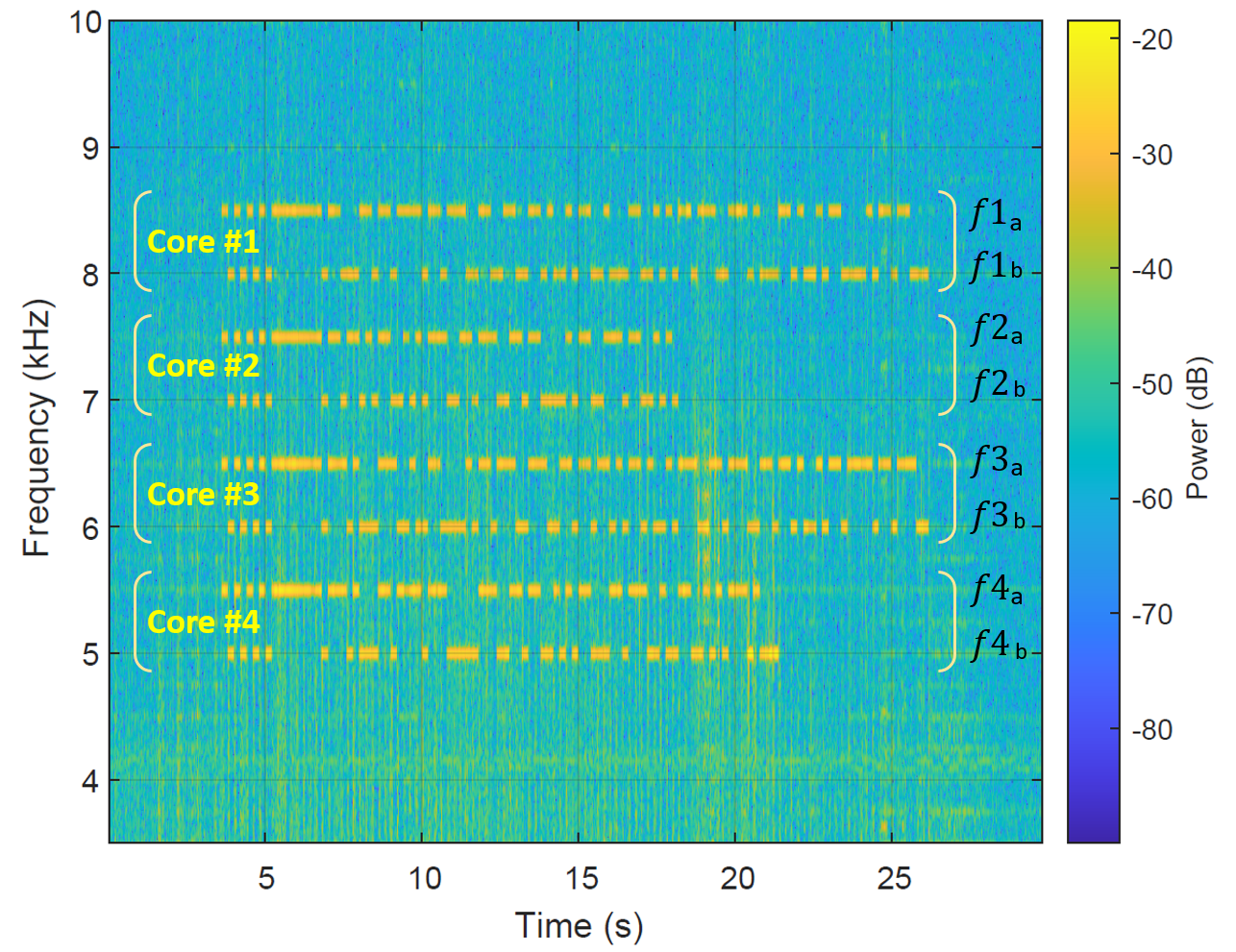}
	\caption{A spectrogram of the multi-carrier modulation with four cores.}
	\label{fig:OFDMSPEC}
\end{figure}

Figure \ref{fig:OFDMSPEC} presents the generation of a signal with four cores. In this case, four threads are bound to four CPU cores. Each thread transmits in a different B-FSK scheme at different switching frequencies ranging from 5 kHz to 8.5 kHz.

\subsection{Transmission Protocol}
Due to the unidirectional nature of this covert channel, acknowledgment and retransmission messages are irrelevant to the transmission protocol. We transmit the data in frame packets composed of calibration bits, payload, and an error detection code.

\begin{figure}
	\centering
	\includegraphics[width=\linewidth]{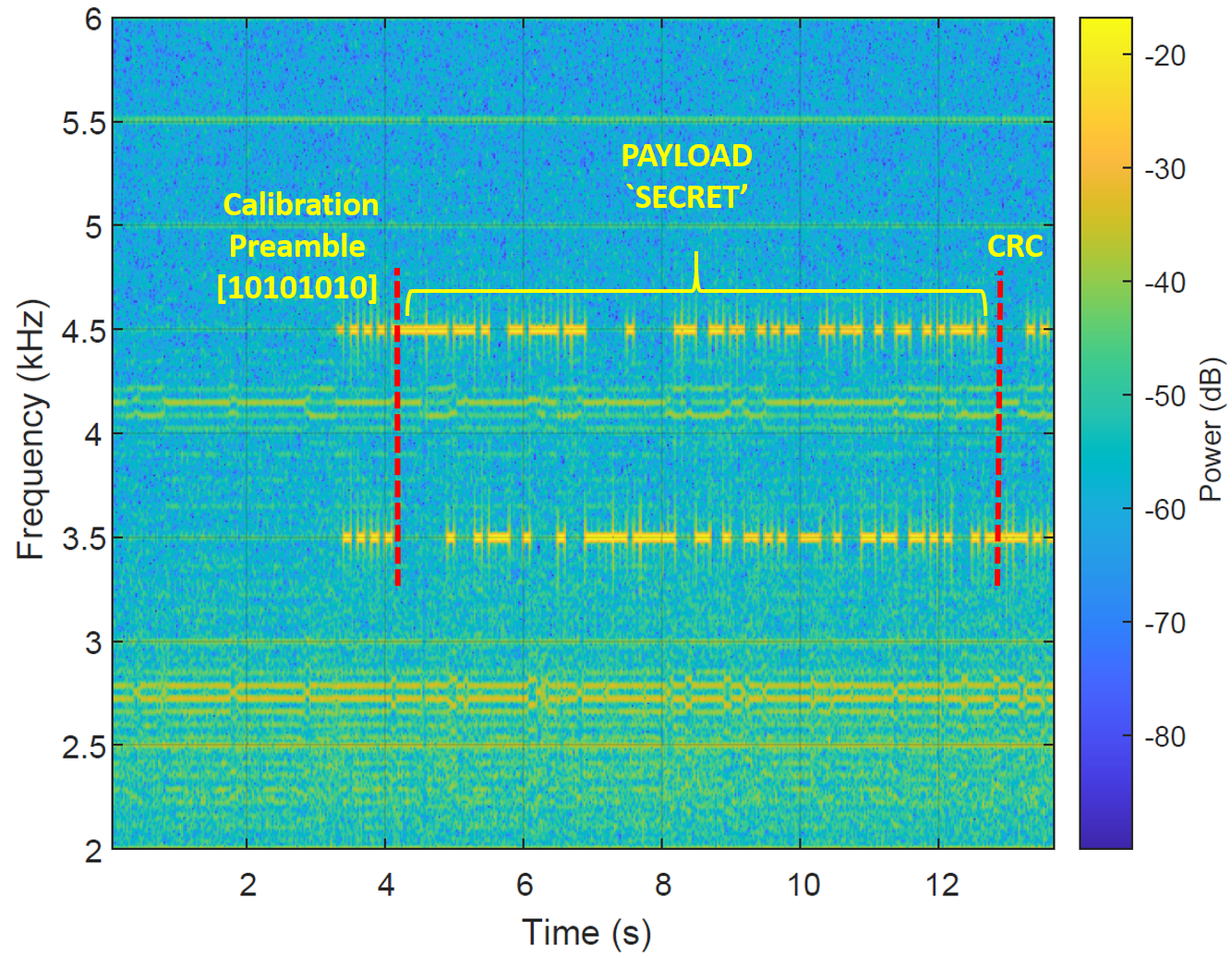}
	\caption{A spectrogram of a frame modulated with B-FSK.}
	\label{fig:frame}
\end{figure}

\begin{itemize}
	\item	\textbf{Calibration bits.} A series of bits transmitted at the beginning of every frame. It consists of eight alternating bits (`10101010'), which help the receiver detect the beginning of a transmission. Since the transmitter parameters are unknown to the receiver in advance, calibration bits are used to extract those parameters (e.g., bit time and amplitude).

	\item \textbf{Payload.} The payload is the raw data to be transmitted. The size can be static or dynamic according to the data transmitted. In the algorithm, we arbitrarily choose 32 bits as the payload size. In the case of dynamic payload size, the first 16 bits encode the size of the actual payload (0-65535 bits) in big-endian.  
	
	\item \textbf{Error detection.} For error detection, an eight-bit CRC (cyclic redundancy check) is added to the end of the frame. The receiver calculates the CRC for the received payload, and an error is detected if it differs from the received CRC bit. 
	
	A spectrogram of a full-frame transmitted from a computer is presented in Figure \ref{fig:frame}. In this case, B-FSK was used for modulation.   
	
\end{itemize}

\section{Receiver Implementation}
\label{sec:rec}
\begin{figure}
	\centering
	\caption{Components of the receiver.}
	\label{fig:demod}
	\includegraphics[width=0.8\linewidth]{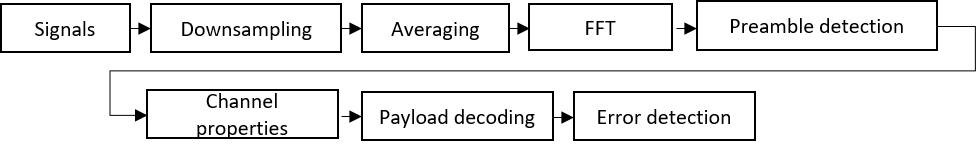}
\end{figure}

A standard audio chip of smartphones, PC, and laptops can sample audio input to 48 kHz. Since we can generate signals in the low-frequency bands of 0 - 48 kHz, we connected a small loop antenna to the 3.5 mm audio input jack and used it as a receiver. A smartphone/laptop antenna can be hidden inside the case or encased as benign headphones or earphones. We implemented two types of receivers: (1) a demodulator in Python running on a laptop with the Linux Ubuntu OS and (2) a demodulator app implemented in Android OS for smartphones. The main functionality of the receiver operates in a dedicated thread. It is responsible for data sampling, signal processing, and data demodulation. The components of the receiver are presented in Figure \ref{fig:demod}. The algorithm receives the data stream, FFT bin size, FFT window size, bit time, and the frequencies $f_{0}$, $f_{1}$ in a case of B-FSK or $f_{0}$, $f_{1}$,...,$f_{m}$ in a case of M-FSK.

\begin{figure}
	\centering
	\includegraphics[width=0.4\linewidth]{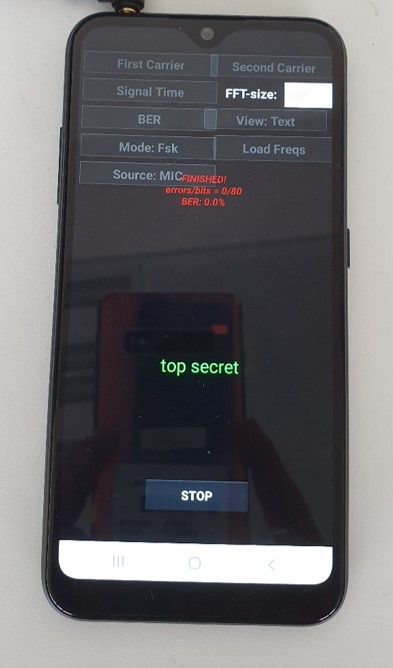}
	\caption{The Android receiver app, decoding the `TOP SECRET' keystrokes exfiltrated from a nearby air-gapped workstation via the COVID-bit covert channel.}
	\label{fig:rec}
\end{figure}

\subsubsection{Signal Processing} The first step is to sample the signal, transform it into the frequency domain, and extract the FSK frequencies. This step applies a fast Fourier transform (FFT) to the array of sampled values. The signal measured is stored in a buffer after using a noise-reduction filter. The raw data from this stage is used later in the demodulation routines. 

\subsubsection{Calibration Detection} The receiver is continuously searching for a calibration sequence to identify a frame header. We used the cross-correlation to detect the preamble as a measure of similarity between signals. If the preamble sequence `10101010' is detected, the state is changed to maintain the demodulation process. Note that after channel characteristics and frequencies have been determined, the raw samples can be down-sampled to the relevant frequency band.

\subsubsection{Demodulation} In this stage, the payload is demodulated and added to the current payload. In the case of dynamic payload size, the size is extracted, and the whole payload is retrieved. When the payload of a certain size and the 8-bit CRC is received, the algorithm returns to the calibration detection state.

Figure \ref{fig:rec} shows the Android receiver app decoding the `TOP SECRET' keystrokes exfiltrated from a nearby air-gapped workstation via the COVID-bit covert channel.

\section{Evaluation}
\label{sec:eval}
We evaluate the covert channel with three types of systems: PC workstations, laptops, and embedded devices. For the evaluation, we measured the electromagnetic signal generated from two types of computers: standard desktop workstations with four and eight physical cores (8 and 16 virtual cores, respectively) and high-end desktop workstations with 12 physical cores (24 virtual cores). We also tested an off-the-shelf laptop computer and a low-power embedded device (Raspberry Pi). We implemented the transmitter as a C program compiled for the Linux Ubuntu OS. The list of computers used for the evaluation and their technical details is provided in Table \ref{table:list}. We implemented a text and file transmitter utility for Ubuntu OS, which transmits text and binary data to the receiver in a sequence of frames, frame after frame. We used two types of receivers listed in Table \ref{table:rec}. A laptop with a demodulator written in Python and a smartphone with a demodulator built as an Android application. For the signal processing and data analysis, we employed the MathWorks MATLAB software.

\begin{table*}[]
	\centering
	\caption{The computers used in the evaluation}
	\label{table:list}
	\resizebox{\textwidth}{!}
	{
		\begin{tabular}{@{}lllll@{}}
			\toprule
			\#     & Type              & Model              & PSU                    & CPU                                                                                                       \\ \midrule
			PC-1   & Standard Workstation        & Silverstone        & FSP300-50HMN 300W      & \begin{tabular}[c]{@{}l@{}}Intel Core i7-4770 CPU @3.4GHz (8 cores) \end{tabular}              \\
			PC-2   & High-end Workstation        & Antec           & PK-450 450W                & 
			Intel Core i9-10920X CPU @3.5GHz (24 cores)
			\\
			PC-3   & Standard Workstation        & Asus           & 450W                & 
			Intel Z490 LGA 1200 ATX i7-10700k (16 cores)
			\\
			Laptop    & Laptop PC & Dell Latitude 7410 & LA65NM190 65W AC Adapter                & Intel(R) Core(TM) i7-6600U CPU @2.60GHz (4 cores)                                                                                                     \\
			IoT    & Embedded/IoT      & Raspberry Pi 3     & DSA-13 PFC-05, 5V 2.5A & \begin{tabular}[c]{@{}l@{}}Quad Core Broadcom BCM2837, 64-bit, ARMv8, Cortex A53\end{tabular} \\ \bottomrule
		\end{tabular}
	}
\end{table*}

\begin{table*}[]
	\caption{The receivers used in the evaluation}
	\label{table:rec}
	\begin{tabular}{@{}lllll@{}}
		\toprule
		\# & Receiver   & Hardware          & Operating System & Sampling rate \\ \midrule
		1  & Laptop     & Dell Latitude 7470, Core i7 Intel(R) Core(TM) 6600-U @2.60GHz (4 cores)                  &  Linux Ubuntu 16.05 LTS                & 0 - 96 kHz    \\
		2  & Smartphone & Samsung Galaxy A01 Octa-core (4x1.95 GHz Cortex-A53 \& 4x1.45 GHz Cortex A53) & Android 10 (Android Q         & 0 - 48 kHz    \\ \bottomrule
	\end{tabular}
\end{table*}

\subsection{Spectral Analysis}
\begin{figure}
	\centering
	\includegraphics[width=0.8\linewidth]{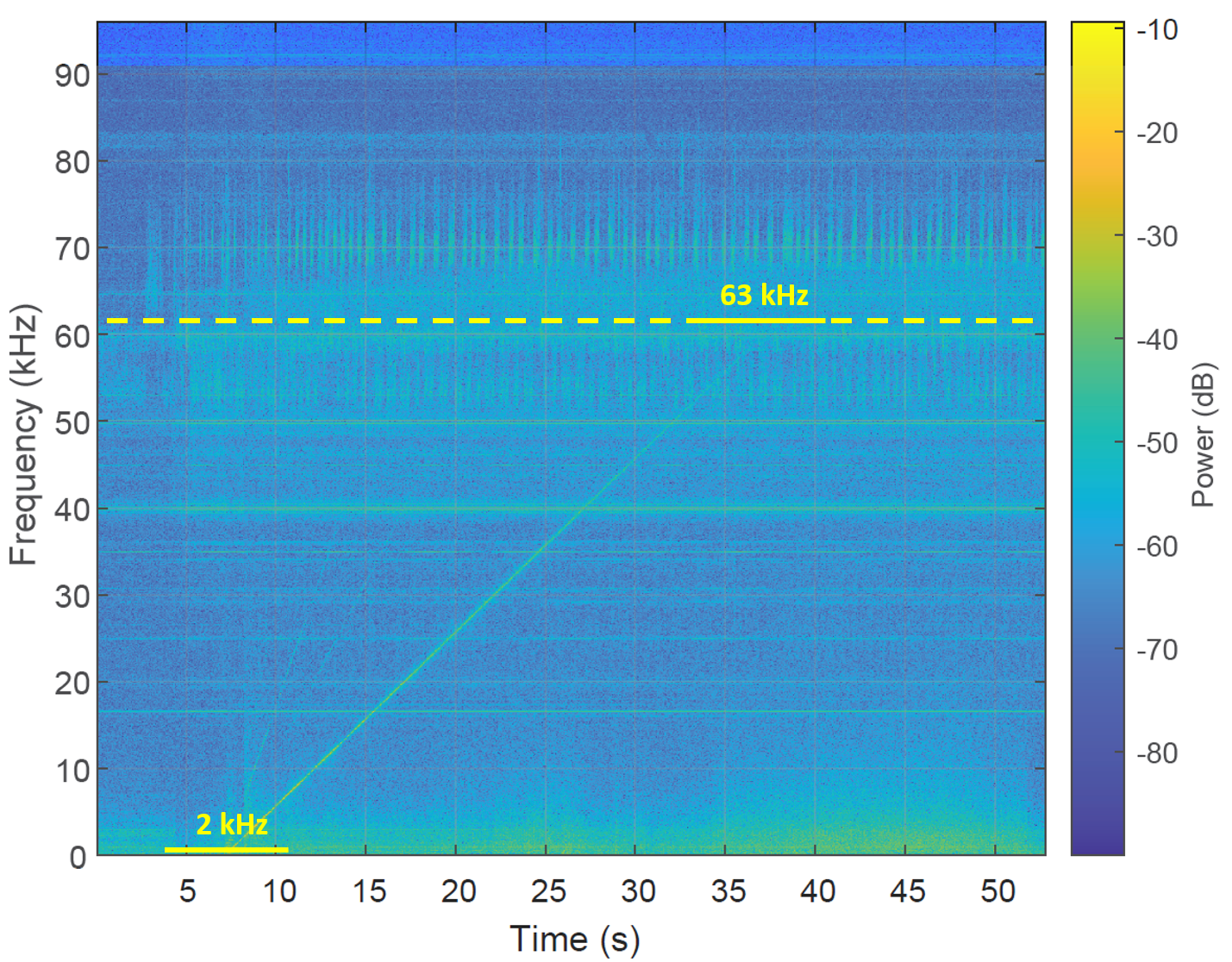}
	\caption{Sweeping signal (`chirp') up to 63 kHz, 100 Hz per step, 1 second per signal, as transmitted from PC-2.}
	\label{fig:sweep}
\end{figure}

We examined the strongest frequency bands generated with the three types of transmitters. Our power spectral density measurements show that the maximum frequency generated in the four transmitters is about 60 kHz, while the strongest signals are bound in the 0 - 30 kHz band. Figure \ref{fig:sweep} shows the spectrogram of a sweeping signal with 100 Hz per step and a step duration of 100 ms, as transmitted from the PC-2. In this case, the whole signal spans from 2 kHz to 63 kHz. The frequency bands measured for the transmitters are listed in Table \ref{tab:bands}. For the experiments, we limited the transmissions to frequency bands of 0 - 24 kHz, which any modern smartphone and laptop receiver can sample.

\begin{table}[]
	\centering
	\caption{The frequency bands}
	\label{tab:bands}
	\begin{tabular}{@{}lll@{}}
		\toprule
		\#     & Band       & Strong band (max)         \\ \midrule
		PC-1   & 2 - 60 kHz & \textless 24 kHz (-30 dB) \\
		PC-2   & 2 - 63 kHz & \textless 30 kHz (-28 dB) \\
		PC-3   & 2 - 59 kHz & \textless 10,30 kHz (-28 dB) \\
		Laptop & 0 - 55 kHz & \textless 40 kHz (-22 dB) \\
		IoT    & 0 - 20 kHz & \textless 20 kHz (-28 dB) \\ \bottomrule
	\end{tabular}
\end{table}

\subsection{Signal-to-Noise Ratio (SNR)}
We measured the Signal-to-Noise (SNR) values for PC-1, PC-2, PC-3, Laptop, and the IoT device. The goal of these measurements is to evaluate the radius of the attack on each device. The results are shown in Table \ref{tab:SNR}. As can be seen, with PC-1, the SNR values decreased from 28.6 dB to 2.5 dB at 60 cm. With PC-2, the effective distance is 200 cm with a signal of 50 dB in the adjacent receiver to 7 dB with a receiver located 200 cm from the computer. With PC-3, the effective distance is 200 cm with a signal of 50 dB in the adjacent receiver to 12 dB with a receiver located 200 cm from the computer. The same distance was achieved with a laptop transmitter, with a signal range of 40 dB to 5 dB at a distance of 100 cm away. An attack on an embedded device (IoT) can be carried out only near the transmitting device.

\begin{figure}
	\centering
	\includegraphics[width=0.7\linewidth]{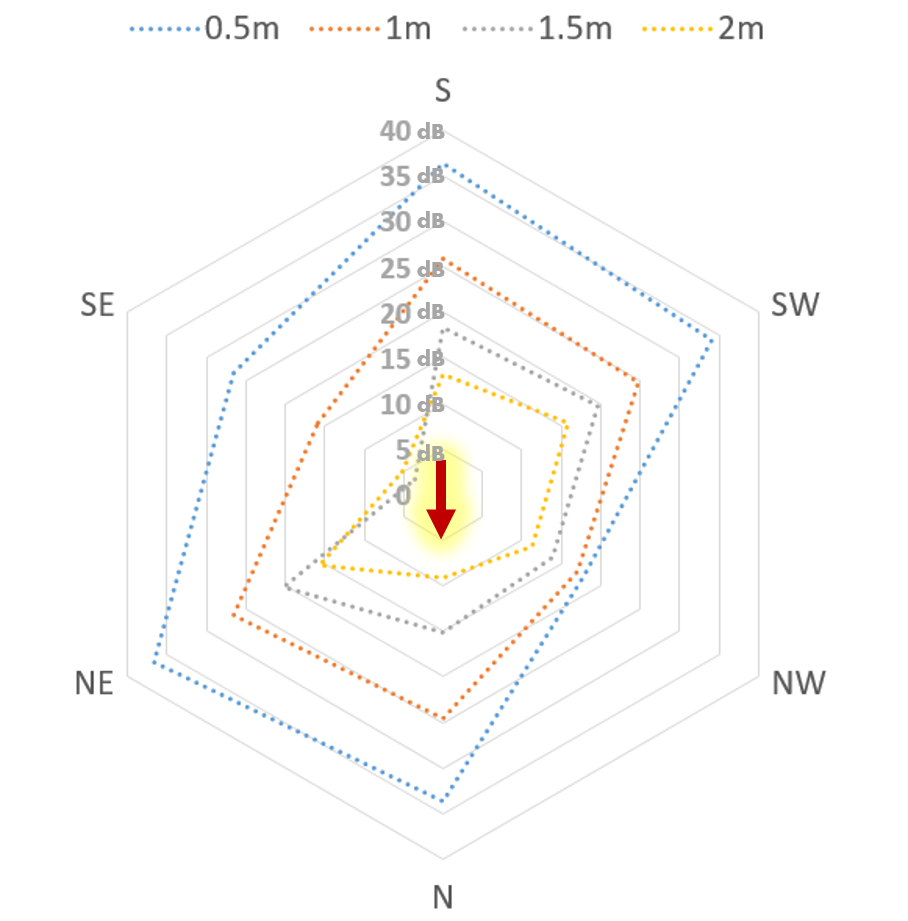}
	\caption{SNR values in six directions for 0.5 m, 1 m, 1.5 m, and 2 m.}
	\label{fig:radar}
\end{figure}

With custom or special antennas, we can reach further distances. Table \ref{tab:SNRt} shows the SNR values measured in different places with various types of antennas, from 0.5 m to 6 m away. Figure \ref{fig:radar} shows the SNR values measured in six directions for 0.5 m, 1 m, 1.5 m, and 2 m.

\begin{table}[]
	\caption{Signal-to-Noise Ratio (SNR) values}
	\label{tab:SNR}
	\begin{tabular}{@{}lllllll@{}}
		\toprule
		\#     & Adjacent & 20 cm   & 40 cm                 & 60 cm                 & 80 cm                 & 100-200 cm                \\ \midrule
		PC-1   & 28.6 dB  & 12.7 dB & 8.8 dB                & 2.5 dB                & \textless 2 dB        & \textless 2 dB        \\
		PC-2   & 50 dB    & 42 dB   & 30 dB                 & 20 dB                 & 17 dB                 & 15 - 7 dB                 \\
		PC-3   & 50 dB    & 40 dB   & 34 dB                 & 32 dB                 & 30 dB                 & 28 - 12 dB                 \\
		Laptop & 40 dB    & 35 dB   & 25 dB                 & 15 dB                 & 10 dB                 & 5 dB                  \\
		IoT    & 35 dB    & 30 dB   & \multicolumn{1}{c}{-} & \multicolumn{1}{c}{-} & \multicolumn{1}{c}{-} & \multicolumn{1}{c}{-} \\ \bottomrule
	\end{tabular}
\end{table}

\begin{table}[]
	\centering
	\caption{Signal-to-Noise Ratio (SNR) values measured at different places with various antennas}
	\label{tab:SNRt}
	\resizebox{\linewidth}{!}{%
		\begin{tabular}{llllllll}
			\hline
			& 0.5 m   & 1 m     & 1.5 m   & 2 m     & 2.5 m  & 3 m    & 6 m  \\ \hline
			SNR & 33.3 dB & 28.9 dB & 24.3 dB & 12.6 dB & 4.7 dB & 5.9 dB & 7.1 dB  \\ \hline
		\end{tabular}%
	}
\end{table}

\subsection{Bit Error Rates (BER)}
In digital communication, the bit error rate (BER) is the number of bits with errors divided by the total number of bits that have
been transmitted. The probability of bit error (BER) approximation of the M-FSK for non-coherent detection in Additive white Gaussian Noise (AWGN) channel is given in equation \ref{eqn4}.

\begin{equation}
	\label{eqn4}
	P_{b,awgn}=\frac{1}{2(M-1)}\sum_{n=1}^{M-1}\frac{(-1)^{n+1}}{n+1}\frac{(M-1)!}{(M-1-n)!n!}e^{-\frac{n(log_{2}M)E_{b}}{(n+1)N_{o}}}     
\end{equation}

We measured the BER for PC-1, PC-2, PC-3, Laptop, and the IoT device at various transmission rates. The results are summarized in Table \ref{tab:BER1} (PC-1 and PC-2), Table \ref{tab:BER2} (Laptop), and Table \ref{tab:BER3} (IoT). With PC-1, PC-2, and  PC-3, we could achieve bit rates of 500 bps with 0\% BER (no errors). For a bit rate of 1000 bps, we maintained BER or 0.8\% for PC-1 and 1\% for PC-2. We maintained bit rates of 100 bps for the laptop transmitter with 0\% BER. For the IoT device, we achieved 200 bps with 0\% BER. Note that the effective transmission rate is decreased with the distance and the environmental noise, as indicated in the SNR measurements. 

\begin{table*}[]
	\centering
	\caption{Bit error rates (BER) for PC-1, PC-2, and PC-3}
	\label{tab:BER1}
	\begin{tabular}{@{}lcccccccc@{}}
		\toprule
		\#   & \multicolumn{1}{l}{5 bps} & \multicolumn{1}{l}{10 bps} & \multicolumn{1}{l}{20 bps} & \multicolumn{1}{l}{50 bps} & \multicolumn{1}{l}{100 bps} & \multicolumn{1}{l}{200 bps} & \multicolumn{1}{l}{500 bps} & \multicolumn{1}{l}{1000 bps} \\ \midrule
		PC-1 & 0\% (no errors)           & 0\% (no errors)            & 0\% (no errors)            & 0\% (no errors)            & 0\% (no errors)             & 0\% (no errors)             & 0.8\%                       & 0.8\%                        \\
		PC-2 & 0\% (no errors)           & 0\% (no errors)            & 0\% (no errors)            & 0\% (no errors)            & 0\% (no errors)             & 0\% (no errors)             & 0.01\%                         & 1.78\%  \\
		PC-3 & 0\% (no errors)           & 0\% (no errors)            & 0\% (no errors)            & 0\% (no errors)            & 0\% (no errors)             & 0\% (no errors)             & 0.05\%                         & 1\%                                        
		\\ \bottomrule
	\end{tabular}
\end{table*}
\begin{table*}[]
		\centering
	\caption{Bit error rates (BER) for the laptop}
	\label{tab:BER2}
	\begin{tabular}{@{}llllll@{}}
		\toprule
		\#     & 5 bps                   & 10 bps                  & 20 bps                  & 50 bps                  & 100 bps                 \\ \midrule
		Laptop & \multicolumn{1}{c}{0\% (no errors)} & \multicolumn{1}{c}{0\% (no errors)} & \multicolumn{1}{c}{0\% (no errors)} & \multicolumn{1}{c}{0\% (no errors)} & \multicolumn{1}{c}{0\% (no errors)} \\ \bottomrule
	\end{tabular}
\end{table*}

\begin{table*}[]
		\centering
	\caption{Bit error rates (BER) for the IoT device}
	\label{tab:BER3}
	\begin{tabular}{@{}lllllll@{}}
		\toprule
		\#  & 5 bps                               & 10 bps                              & 20 bps                              & 50 bps                              & 100 bps                             & 200 bps                             \\ \midrule
		IoT & \multicolumn{1}{c}{0\% (no errors)} & \multicolumn{1}{c}{0\% (no errors)} & \multicolumn{1}{c}{0\% (no errors)} & \multicolumn{1}{c}{0\% (no errors)} & \multicolumn{1}{c}{0\% (no errors)} & \multicolumn{1}{c}{0\% (no errors)} \\ \bottomrule
	\end{tabular}
\end{table*}

\subsection{Number Of Cores}
\begin{figure}
	\centering
	\includegraphics[width=0.6\linewidth]{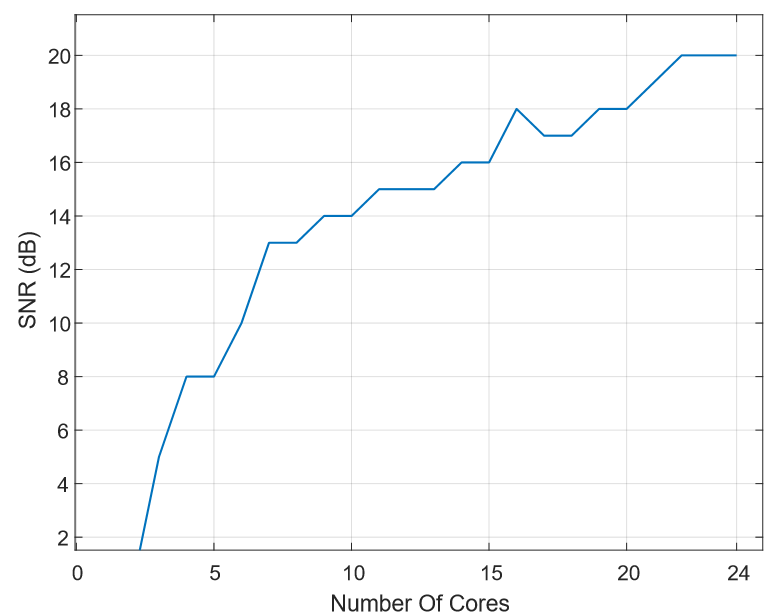}
	\caption{SNR given number of transmitting cores in PC-2.}
	\label{fig:PCSNR24}
\end{figure}

The number of cores that participate in transmission directly affects the energy consumption in the system. Consequentially, the generated signal is stronger with a higher number of transmitting cores. Figures \ref{fig:PCSNR24} emphasizes this phenomenon for PC-2, increasing signal strength from 2 dB with two cores to 20 dB with 24 cores participating in the transmission. Notably, the effect is more limited in the laptop transmitter, with a slight increase of only 2 dB between two and eight transmitting cores. This is due to the limited dynamics in the laptop's energy consumption due to its power management and power-saving profiles.      

\begin{table}[]
		\centering
	\caption{Measurements of the signal with background workloads}
	\label{tabs:workloads1}
	\begin{tabular}{@{}llcc@{}}
		\toprule
		\# & Workload           & \multicolumn{1}{l}{Laptop} & \multicolumn{1}{l}{PC-2} \\ \midrule
		1  & No workload (Idle) & 39 dB                      & 30 dB                    \\
		2  & Word processor     & 35 dB                      & 30 dB                    \\
		3  & Files copy (I/O)   & 35 dB                      & 30 dB                    \\
		4  & Video playing      & 33 dB                      & 30 dB                    \\
		5  & Calculations (CPU) & 34 dB                      & 30 dB                    \\ \bottomrule
	\end{tabular}
\end{table}

\subsection{Background Processes}
\begin{figure}
	\centering
	\caption{Signal generated with various background workloads.}
	\label{fig:workload}
	\includegraphics[width=0.8\linewidth]{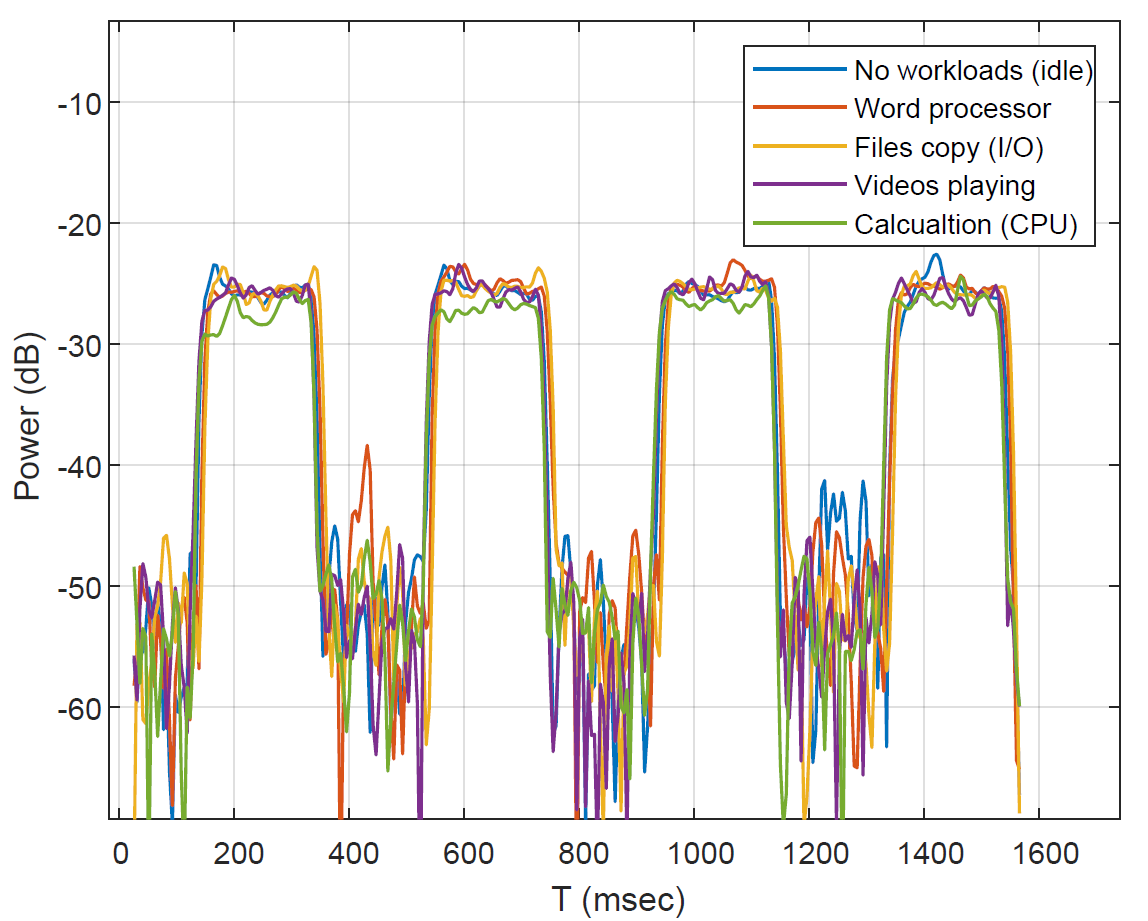}
\end{figure}

In addition to transmitting process activity, other processes can affect the momentary power consumption of the system. We measured the effect of other processes in the system on the quality of the signal. We measured five types of workloads; (1) Idle system, where no other special workloads are running concurrently to the signal generation, (2) running a word processor application, (3) executing I/O operations where a batch of files are copied from one location to another using the Linux cp command, (4) playing a video with the VLC media player, and (5) performing intensive CPU calculations. The results are depicted in Figure \ref{fig:workload} and detailed in Table \ref{tabs:workloads1} for laptop and PC-2 transmitters. As can be seen, there is a neglected effect on the signal for all five workloads, including I/O and CPU workloads. With PC-2, we measured no effect on the quality of the generated signal. This is due to the number of cores in the system available to execute the additional workload and the efficient direct memory access (DMA) employed on this high-end modern motherboard.   

\subsection{Data Transmissions}
Table \ref{tab:data} present the time it takes to exfiltrate various types of information. With keystrokes encoded with 6 bits per key, keylogging is exfiltrated in real-time at almost all speeds. A brief amount of information, such as IP \& MAC addresses, takes between 16 seconds in the slowest (5 bps) to less than 0.1 seconds in the fastest (1000 bps). A Bitcoin private key (AES 256) can be exfiltrated in between 51.2 seconds and less than one second depending on the speed. Depending on the speeds, more significant amounts of data, such as the 4096-bits RSA encryption key, may take ten minutes to four seconds. A large amount of information (e.g., one hour of keylogging and text files) is relevant mainly for 50 bps and higher transmission speeds.  

\begin{table}[]
	\caption{Transmission times}
	\label{tab:data}
	\resizebox{\columnwidth}{!}{
		\begin{tabular}{@{}lccccc@{}}
			\toprule
			Information                        & \multicolumn{1}{l}{Size (bits)} & \multicolumn{1}{l}{5 bps} & \multicolumn{1}{l}{50 bps} & \multicolumn{1}{l}{100 bps} & \multicolumn{1}{l}{1000 bps} \\ \midrule
			Keylogging                           & 6 bits/key                         & near real-time                         & real-time                        & real-time                          & real-time                 \\
			IP / MAC                           & 80 bits                         & 16                        & 1.6                        & 0.8                         & \textless 0.1                \\
			Keylogging (1 hour)                & $\sim$20k                       & 4000                      & 400                        & 200                         & 20                           \\
			Bitcoin private key                            & 256                             & 51.2                      & 5.12                       & 2.56                        & 0.256                        \\
			RSA Encryption Key (4096)          & 4096                            & 819.2                     & 81.92                      & 40.96                       & 4.096                        \\
			Small file (10K)             & 80k                             & 16000                     & 1600                       & 800                         & 80                           \\
			Smartcard Credentials  & 300                             & 60                        & 6                          & 3                           & 0.3                          \\ \bottomrule
	\end{tabular}}
\end{table}

\subsection{Attenuation}
As described in the attack model, the attacker might be an insider or malicious visitor carrying his mobile phone or another type of receiver in the vicinity of the air-gap computer. The malicious person doesn't have physical access to the restricted room of the air-gap computer but only to an external, less classified area, e.g., a corridor outside the room. We measured the effect of a typical wall on the electromagnetic field in specific frequency bands. Attenuation is measured in decibels and varies with the frequency of the RF signal source as well as the type and density of materials. 

\begin{figure}
	\centering
	\caption{Measurements of electromagnetic field attenuation in open space and behind a wall.}
	\label{fig:wall}
	\includegraphics[width=0.8\linewidth]{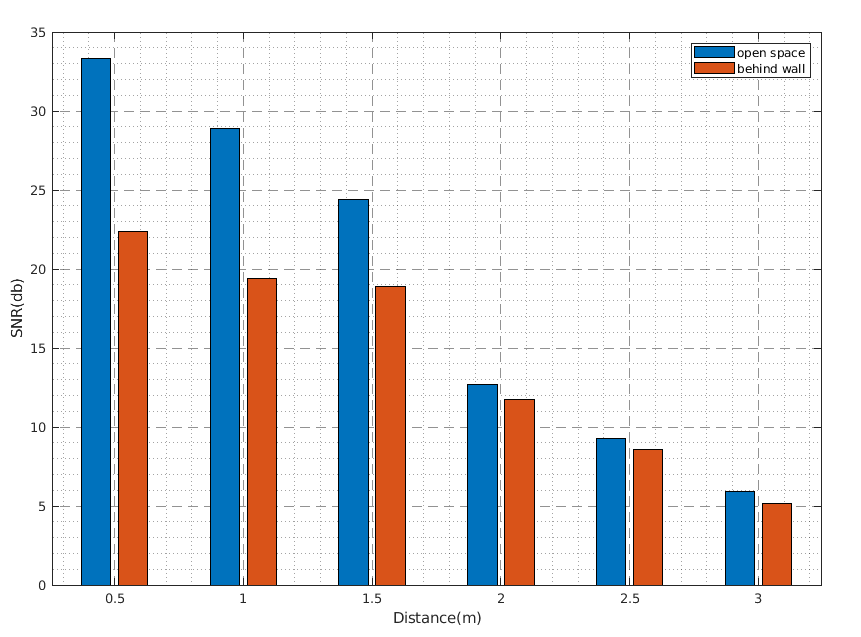}
\end{figure}

Figure \ref{fig:wall} shows the measurements of the PC-3 transmission in open space and behind a concrete wall 25 cm thick in distances of 0.5 m to 3 m.

\subsection{Virtualization (VM)}
\begin{figure}
	\centering
	\includegraphics[width=0.8\linewidth]{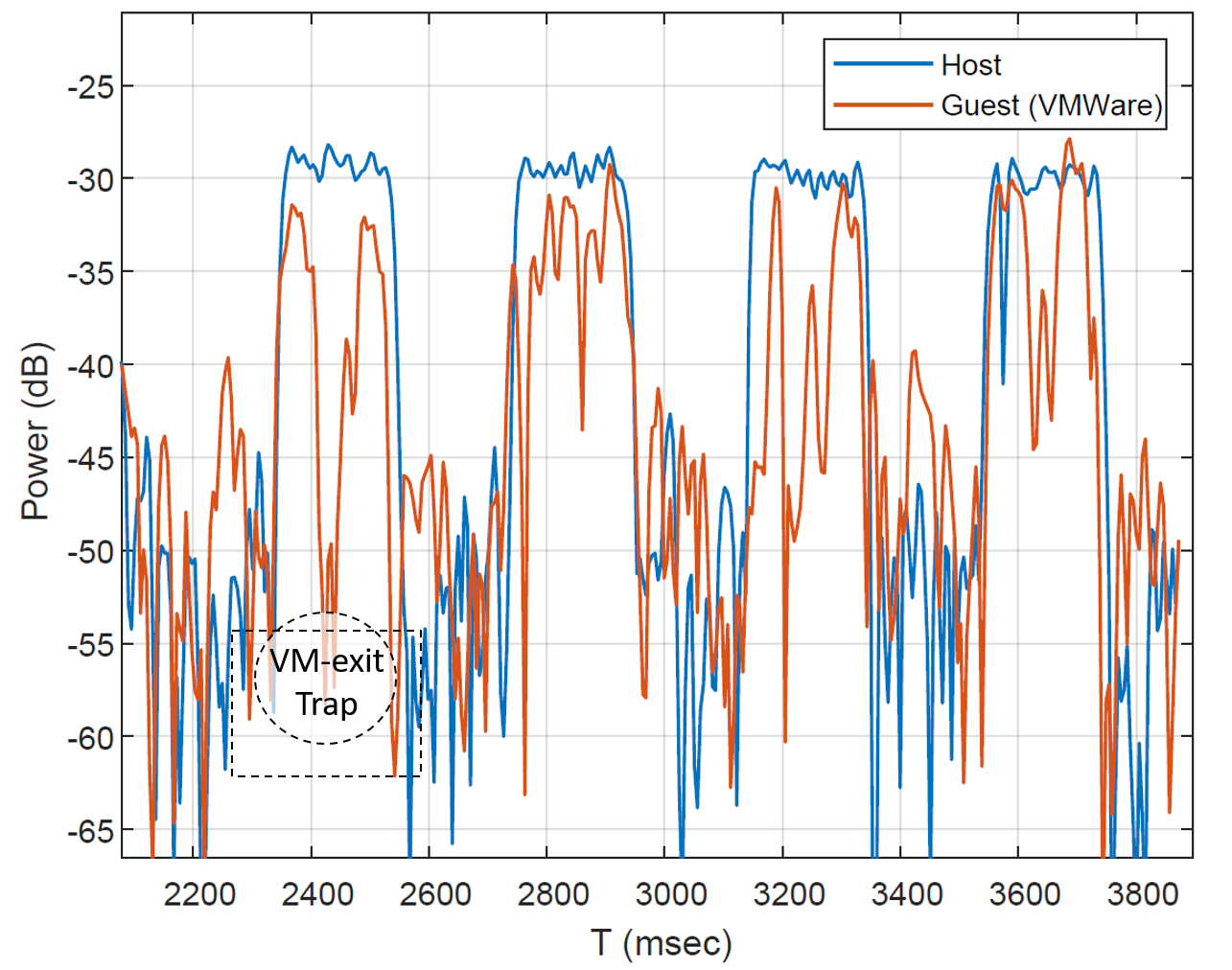}
	\caption{The effect of a virtualization layer (hypervisor) on the signal.}
	\label{fig:vmware}
\end{figure}

Virtual machines (VM) are used in many modern computing environments, including personal workstations, on-premise servers, high-performance computing (HPC), and large cloud farms. One main advantage of virtualization is isolation from the physical host via hardware abstraction layers. The Virtual Machine Monitor (VMM), or hypervisor, intercepts the privileged instructions, memory accesses, and I/O operations and handles them according to the guest virtual machine (VM) permissions. We examined how the execution of the transmitting threads within a VM affects the output signal. Figure \ref{fig:vmware} shows the waveform of an alternating sequence signal generated from a host machine (bare metal) and a guest virtual machine. We used VMware Workstation 16.2 for the tests, which runs Linux Ubuntu 16.04.7 LTS. The experiments show that the signals generated from the VM may include one to two short interruptions. This is due to the operation of VM-exit traps to the hypervisor handler, which disrupts the execution of the signal generation code on the CPU. We found that the signal degradation range from two to eight decibels (2 dB - 8 dB).

\section{Countermeasures}
\label{sec:counter}
Several defensive and protective countermeasures can be taken to defend against the proposed covert channel. 
As discussed in Section \ref{sec:tech} the electromagnetic signals are generated by careful orchestration of the load on the CPU cores using one or more threads. Security systems such as malware protection and detection applications can monitor how running threads utilize the CPU cores to detect suspicious patterns. In the case of COVID-bit, threads that persistently change the CPU utilization would be reported for further forensic investigation. However, there are two main challenges with such a detection approach. First, since many legitimate applications are CPU-bound and perform various workloads at different patterns, the system may be prone to a high rate of false positives. Second, tracing the activities of threads in the operating system with the fine-grained resolution is not a trivial task since the signal generation compound of non-privileged CPU instructions such as loops and calculations. Monitoring such instructions at runtime necessitates entering a trap mode, which severely degrades performance \cite{guri2015gsmem}. Applying static and dynamic analysis to the thread code (.text section) before execution is possible, but it also suffers from an inherent weakness in that it can easily be bypassed by malware with evasion techniques and also has runtime overhead \cite{damodaran2017comparison}. Notably, previous work suggested using opcode level analysis at runtime to detect malicious behavior. Carlin et al. suggested using dynamic opcode analysis to detect cryptominers and other threats \cite{carlin2018detecting}.

\begin{figure}
	\centering
	\caption{The effect of frequency locking.}
	\label{fig:fl}
	\includegraphics[width=0.6\linewidth]{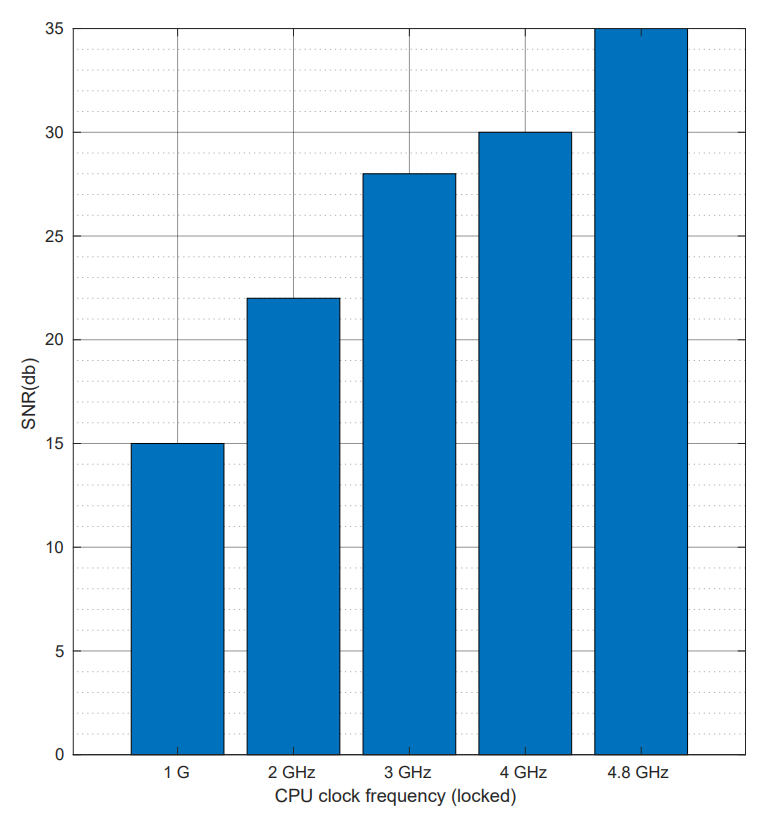}
\end{figure}

Another preventive measure that may be taken is to fixate the CPU on a certain frequency, avoiding dynamic frequency scaling. This can be done using the UEFI configuration or operating system level configuration tools \cite{UbuntuMa84:online}. However, our experiments showed that locking the CPU frequency has only a limited effect on the signal and does not prevent it entirely. This is because locking the frequency determines the whole energy of the carrier wave but does not prevent the effect of switching on the low-frequency sidebands. Figure \ref{fig:fl} shows the impact of frequency locking on the signal. Setting the core frequency to 4.8 GHz, 4 GHz, 3 GHz, 2 GHz, and 1 GHz yields SNR of 35 dB, 30 dB, 28 dB, 22 dB, and 15 dB, respectively. While this defensive approach technically reduces the signal and the effective distance, it limits the energy savings of many system components, including CPU cores. This makes this solution less usable. 

\begin{figure}
	\centering
	\caption{The signal without jamming (A), 16 jamming threads (A1), 32 jamming threads (A2), and 48 jamming threads (A3).}
	\label{fig:jam}
	\includegraphics[width=1\linewidth]{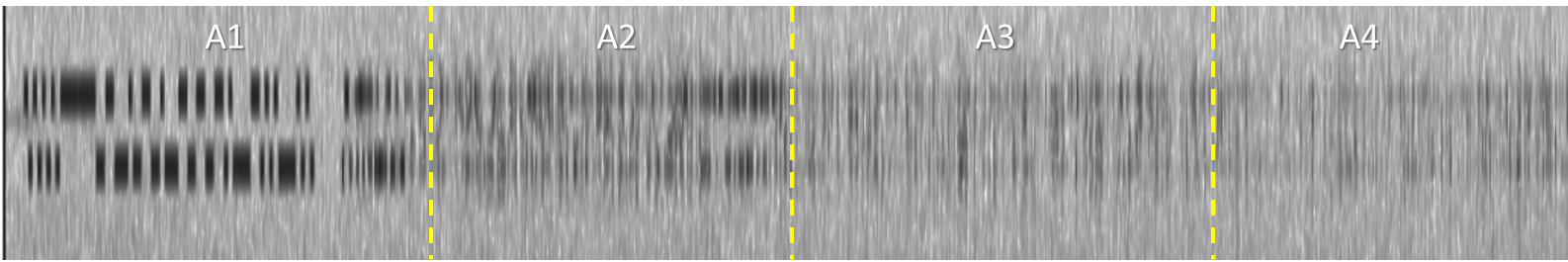}
\end{figure}

To disrupt the generation of a signal, it is possible to initiate random workloads on the CPU processors whenever suspicious activity is detected. The random workload yields variations in the switching frequency and interferes with the malicious process's carrier. Our tests show that a single process bound to the same core as the transmitter can significantly reduce the SNR of the original signal. Figure \ref{fig:jam} shows the spectrogram of signal without jamming (A) with 37 dB, 16 jamming threads (A1) with 7.5 dB, 32 jamming threads (A2) with 2.5 dB, and 48 jamming threads (A3) with 1.5 dB.

However, this approach has an inherent overhead of the extra CPU utilization and can also be evaded by malicious code. One solution might be injecting the jammer code directly from the kernel level. However, even kernel-level code can be compromised and manipulated by advanced threats such as kernel exploits and rootkits \cite{blunden2012rootkit}\cite{tian2019kernel}. 

In our case, hardware countermeasures may include external electromagnetic receivers which persistently monitor the relevant spectrum of 0 - 60 kHz. Note that this solution is considered a \textit{trusted} since it operates outside the compromised system. Another hardware-based solution is to use a radio frequency jamming system to block the reception or transmission of signals in the relevant frequency bands of 0 - 60 kHz. Certain radio jamming equipment is restricted or illegal in many countries (including Europe and the U.S.) and can not be used in all scenarios. Regulation-based countermeasures may include using the TEMPEST standards to prohibit or monitor the existence of radio receivers in the vicinity of classified or sensitive systems (e.g., NSTISSAM TEMPEST/2-95 \cite{NSTISSAM75:online}). In our case, recording devices, such as mobile phones, should be banned from the area of air-gapped systems. Note that such solutions are less relevant to the adversarial attack model in which malicious insiders and implanted hardware are involved. Physical security may be using a Faraday cage or Faraday shielding to prevent electromagnetic emissions from the computer systems. This approach involves encasing the entire system in a thin layer of conductive material or metal mesh. This approach is costly and impractical as a large-scale solution. Prior work showed that even Faraday shields could be bypassed using magnetic-based covert channels. Several standards measure and limit the radiated and conducted emission a given system generates and can tolerate without malfunction. For example,  CISPR11/EN55011 are defined for industrial, medical, and scientific applications, CISPR13/EN55013 for consumer applications, CISPR14/EN55014 for home appliances and power tools, CISPR15/EN55015 for lighting equipment, and CISPR22/EN55022 for computing applications \cite{1008Aug224:online}. 
The defensive countermeasures and their weaknesses are listed in Table \ref{tab:countermeasures}.

\begin{table}[]
	\caption{Countermeasures}
	\label{tab:countermeasures}
	\resizebox{\columnwidth}{!}{
		\begin{tabular}{@{}ll@{}}
			\toprule
			Countermeasure                           & Weaknesses                               \\ \midrule
			Runtime processor monitoring             & Trace overhead; false positives         \\
			Frequency locking            & Energy waste         \\
			Signal disruption (workload injection)  & Runtime overhead; evasion techniques     \\
			Electromagnetic monitoring               & False positives; deployment; cost        \\
			Regulations (e.g., mobile phone banning) & Malicious insiders; supply-chain attacks \\
			Faraday shielded systems                 & Cost; deployment                         \\ \bottomrule
	\end{tabular}}
\end{table}

\section{Conclusion}
\label{sec:conclusion}
This paper introduced COVID-bit, a new type of attack that enables the leaking of sensitive information from air-gapped systems to nearby receivers carried by a malicious person. We showed that malware on air-gapped computers could generate electromagnetic radiation in specific low-frequency bands. The malicious code exploits modern computers' dynamic power and voltage regulation and manipulates the loads on CPU cores. Sensitive information can be encoded in low-frequency electromagnetic radiation and intercepted by a nearby receiver (e.g., a smartphone). We showed that this technique works from a normal user-level process and operates within a virtual machine (VM). We implemented a transmitter of text and binary files for air-gapped computers. We presented signal generation, data modulation, and demodulation and the experimental and evaluation results. With COVID-bit, we successfully leaked information from air-gapped workstations to nearby receivers at a bandwidth of 1000 bits/sec.

\def\UrlBreaks{\do\/\do-}
\balance
\bibliographystyle{plain}
\bibliography{../../../airgapnetworks,switch,../../../AirGap,../../../AirGapCases,../../../mobile,../../../AirGapTools,../../../PowerSupply,../../../magnetic}

\end{document}